\documentclass[reprint, amsmath]{revtex4-2}
\usepackage{graphicx}
\usepackage{hyperref}
\usepackage{xcolor}
\usepackage{subcaption}

\graphicspath{{./figures/}}

\DeclareUnicodeCharacter{2009}{\,}

\begin{document}

\title{Short-range order and compositional phase stability in refractory high-entropy alloys via first principles theory and atomistic modelling: NbMoTa, NbMoTaW and VNbMoTaW}

\author{Christopher D. Woodgate}
\email[]{C.Woodgate@warwick.ac.uk}
\affiliation{Department of Physics, University of Warwick, Coventry, CV4 7AL, United Kingdom}
\author{Julie B. Staunton}
\email[]{J.B.Staunton@warwick.ac.uk}
\affiliation{Department of Physics, University of Warwick, Coventry, CV4 7AL, United Kingdom}

\begin{abstract}
Using an all-electron, first principles, Landau-type theory, we study the nature of short-range order and compositional phase stability in equiatomic refractory high entropy alloys, NbMoTa, NbMoTaW, and VNbMoTaW. We also investigate selected binary subsystems to provide insight into the physical mechanisms driving order. Our approach examines the short-range order of the solid solutions directly, infers disorder/order transitions, and also extracts parameters suitable for atomistic modelling of diffusional phase transformations. We find a hierarchy of relationships between the chemical species in these materials which promote ordering tendencies. The most dominant is a relative atomic size difference between the $3d$ element, V, and the other $4d$ and $5d$ elements which drives a B32-like order. For systems where V is not present, ordering is dominated by the difference in filling of valence states; pairs of elements which are isoelectronic remain weakly correlated to low temperatures, while pairs with a valence difference present B2-like order. Our estimated order-disorder transition temperature in VNbMoTaW is sufficiently high for us to suggest that SRO in this material may be experimentally observable.
\end{abstract}

\maketitle

\section{Introduction}

A recent development in the field of materials science is the discovery  of the so-called high-entropy alloys (HEAs)\cite{cantor_microstructural_2004, yeh_nanostructured_2004, gao_high-entropy_2016, george_high-entropy_2019}, of which the first examples were synthesised by Cantor {\it et al.} \cite{cantor_microstructural_2004} and Yeh {\it et al.} \cite{yeh_nanostructured_2004}. Yeh {\it et al.} attributed the stabilisation of the single phase solid solution to the large contribution to the free energy of the system from the configurational entropy, hence the term `high-entropy'. These systems are also referred to as `multicomponent' or `multi-principal element' alloys, and systems with three (or sometimes four) elements are occasionally referred to as `medium entropy'. They are metallic alloys in which three or more elements are combined in roughly equal proportions to form a single phase solid solution, with a simple underlying lattice structure. Although these systems possess compositional disorder, it is known from both theory and experiment that atoms in these materials do not arrange themselves entirely randomly, and a degree of atomic short-range order (SRO) is both theoretically predicted and has been experimentally observed\cite{singh_atomic_2015, fernandez-caballero_short-range_2017, zhang_local_2017, schonfeld_local_2019, chen_direct_2021, zhang_short-range_2020}. 

It is understood that SRO affects material properties\cite{xing_short-range_2022, wang_chemical_2021, li_complex_2020, zhang_short-range_2020, yin_yield_2020, ding_tunable_2018}, and therefore a key challenge for computational modellers and theorists is not only to predict what multicomponent single phase alloys can form but also understand the nature of SRO in those that do. This should assist in the design of new HEAs and also assist in the production of these materials for applications by guiding the annealing process and suggesting required temperatures to either promote or impede the development of SRO. The space of candidate HEAs alloys is vast and, therefore, techniques which are computationally expensive or which scale poorly with increasing number of chemical species should be used sparingly. Modelling techniques are needed which are  both accurate and computationally efficient.

A family of techniques which satisfies these requirements and is arguably well suited to modelling HEAs and their inherent disorder at scale contains effective medium theories\cite{tian_alloying_2017,niu_first_2016,robarts_extreme_2020}, such as the coherent potential approximation (CPA)\cite{soven_coherent-potential_1967}. Typically, to examine compositional order, such approaches, combined with density functional theory (DFT),  seek to analyse the energetic favorability of particular chemical fluctuations when instigated in high-temperature, disordered solid solution phases using the concentration wave formalism\cite{singh_atomic_2015, schonfeld_local_2019, khachaturyan_ordering_1978, gyorffy_concentration_1983, staunton_compositional_1994, singh_tuning_2019, singh_first-principles_2020,singh_martensitic_2021}.

Other modelling techniques have also been used to study compositional phase behaviour in HEAs. These include large scale super-cell calculations with energies evaluated via DFT, molecular dynamics simulations based on interatomic potentials, CALPHAD and semi-empirical calculations\cite{ferrari_frontiers_2020,feng_design_2016,feng_phase_2018,widom_modeling_2018,sorkin_generalized_2020,ikeda_ab_2019, troparevsky_criteria_2015, feng_first-principles_2017, gao_thermodynamics_2017, gorsse_about_2018}. Machine learning approaches and cluster expansions have been applied to develop Hamiltonians with which to perform atomistic modelling\cite{pei_machine-learning_2020, huang_machine-learning_2019,fernandez-caballero_short-range_2017,liu_monte_2021}. Supercell calculations on HEAs with energies evaluated via DFT are limited to a subset of possible configurations, even when studying relatively small supercells, because of the high computational cost of such calculations. A more complete exploration of the phase space is provided by atomistic models, but the parameters used as inputs to these models come from a range of origins, and the underlying physics driving ordering is not always explored.

\begin{figure*}[t]
\centering
\begin{subfigure}{0.32\textwidth}
\centering
\includegraphics[width=\textwidth]{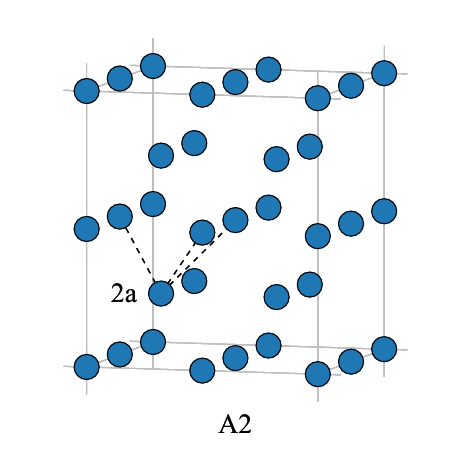}
\end{subfigure}
\begin{subfigure}{0.32\textwidth}
\centering
\includegraphics[width=\textwidth]{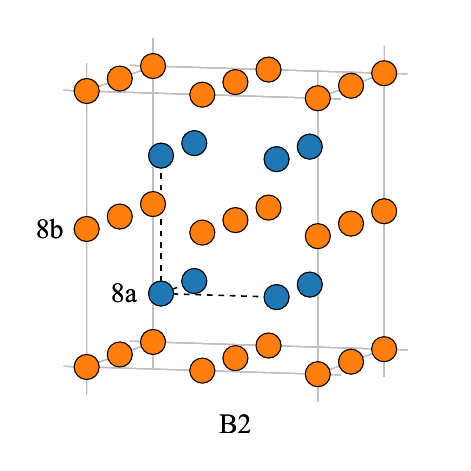}
\end{subfigure}
\begin{subfigure}{0.32\textwidth}
\centering
\includegraphics[width=\textwidth]{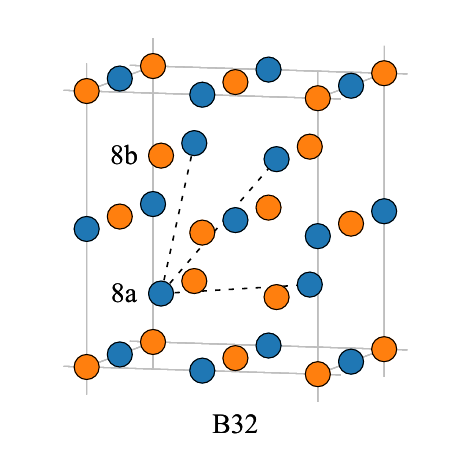}
\end{subfigure}
\caption{Examples of ordered structures based on the bcc lattice which can be described by concentration waves. $A2$ represents a disordered alloy. B2 can be described by modes $\mathbf{k}=\{0,0,1\}$, while B32 is described by modes $\mathbf{k}=\left\{\frac{1}{2},\frac{1}{2},\frac{1}{2}\right\}$}
\label{fig:concentration_waves}
\end{figure*}

In an earlier work, we outlined our approach to studying compositional order in HEAs with an {\it ab initio} electronic structure model and applied it to the Ni-based, face-centred cubic (fcc) Cantor-Wu systems\cite{khan_statistical_2016, woodgate_compositional_2022}, arguably the prototypical HEAs. We were able to demonstrate that interactions in those systems extended beyond nearest-neighbour distance and also that interactions were poorly approximated as pseudobinary. The Cantor-Wu systems are well-studied both computationally and experimentally, and our method gave good agreement with existing literature on SRO in these systems for minimal computational cost.
We now turn our attention to  another well-studied set of high-entropy materials, the refractory HEAs, which form on a body-centred cubic (bcc) lattice. Originally synthesised by Senkov {\it et al.} \cite{senkov_refractory_2010, senkov_mechanical_2011}, these materials possess extraordinary physical properties on account of the base elements from which they are constructed, comparable to or even superior than the Ni-based HEAs, and are therefore good candidates for next-generation engineering materials, particularly for high-temperature fission and fusion applications\cite{fernandez-caballero_short-range_2017, zou_ultrastrong_2015, zou_size-dependent_2014}.

We choose to study the two original equiatomic refractory HEAs, NbMoTaW and VNbMoTaW, along with the medium-entropy, equiatomic NbMoTa. This represents a series of a three-component, four-component, and five-component equiatomic alloys with increasing configurational entropy, all of which are known to form single-phase solid solutions \cite{senkov_refractory_2010, senkov_mechanical_2011, li_complex_2020}. Although the literature on compositional order in these systems is not as large as for the Cantor-Wu alloys, a number of earlier works have highlighted some interesting behaviors and suggest these systems warrant further study \cite{huhn_prediction_2013, fernandez-caballero_short-range_2017, singh_ta-nb-mo-w_2018, widom_hybrid_2014, kormann_interplay_2016, kormann_long-ranged_2017}. 

Our aim is not just to describe the nature of compositional order in these specific materials, but also to elucidate its origins in terms of the materials' electronic structure and obtain physical insight into the origins of compositional order stability to aid material design. To that end, we study a number of binary subsystems within the same formalism as used for the multicomponent systems to extract qualitatively the mechanisms driving ordering in alloys consisting of refractory metals.
For the materials chosen, we provide a complete description of the nature of SRO and the temperatures at which it emerges. We provide insight into its origins in terms of the electronic structure of the solid-solution. We also provide pairwise interchange parameters suitable for further atomistic modelling on these systems which we demonstrate.

This paper is laid out as follows. First, in \ref{sec:theory}, we outline briefly the underlying theory and our methodology for studing compositional order in multicomponent alloys. Then, in section \ref{sec:results}, we provide results from electronic structure calculations for the solid solutions, linear response analysis of atomic SRO, and atomistic modelling of the phase stability for the three considered systems. Rather than just providing predictions for the nature of compositional order in these materials, we also use details of the materials' electronic structure to give qualitative insight into its origins, transferable to other multi-component alloys. Finally, in \ref{sec:conclusions}, we summarise our results and give an outlook on further work.

\section{Theory}
\label{sec:theory}

Our technique for modelling compositional order in multicomponent alloys differs from many alternative techniques and uses a Landau-type expansion of the free energy of the system to obtain the two-point correlation function, a SRO parameter, {\it ab initio}.  Effects on the electronic structure and the rearrangement of charge, { in response to an applied inhomogeneous chemical perturbation}, are fully included \cite{khan_statistical_2016, woodgate_compositional_2022}. This is an extension of the $S^{(2)}$ theory developed for binary alloys \cite{gyorffy_concentration_1983, staunton_compositional_1994}. Our calculations assume a fixed ideal lattice, bcc for the studies in this paper, which represent the averaged positions of the atomic positions in the multicomponent solid solutions and is a major reason for the low computational cost.  For an alloy, descriptors for the modelling of ``small'' atoms, like V, mixing with ``big'' atoms, like Ta, turn out to be effective charge transfers to the small from the big atoms, screened by the valence electrons.

The theory has its groundings in statistical physics and the seminal papers on concentration waves by Khachaturyan~\cite{khachaturyan_ordering_1978} and Gyorffy and Stocks ~\cite{gyorffy_concentration_1983}. A substitutional alloy is described by a set of site-wise occupancies, $\left\{\xi_{i\alpha}\right\}$, where $\xi_{i\alpha}=1$ if site $i$ contains an atom of species $\alpha$ and $\xi_{i\alpha}=0$ otherwise. We denote the average value of $\xi_{i\alpha}$ by 
\begin{equation}
c_{i\alpha}  \equiv \langle \xi_{i\alpha} \rangle   
\end{equation}
and the value of the total overall concentration of species $\alpha$ is given by $c_\alpha = \frac{1}{N} \sum_i c_{i\alpha}$, where $N$ is the number of lattice sites.
Above any order/disorder transition temperature, in the solid solution phase, the $\left\{c_{i\alpha}\right\}$ will be spatially homogeneous and each will simply take the value of the total overall concentration of species $\alpha$, $c_\alpha$. Below an order/disorder transition temperature they will acquire a spatial dependence as the translational symmetry is broken. These are therefore our long-range order (LRO) parameters. It is most convenient to lattice Fourier transform and describe ordering in reciprocal space via concentration waves, $c_\alpha(\mathbf{k})$~\cite{khachaturyan_ordering_1978, gyorffy_concentration_1983}. An ordered structure can then be represented by
\begin{equation}
c_{i\alpha} = c_\alpha + \sum_{\mathbf{k}} e^{i \mathbf{k} \cdot \mathbf{R}_i} c_\alpha(\mathbf{k}).
\end{equation}
Examples of some binary ordered structures on the bcc lattice and the concentration waves describing them are given in Figure~\ref{fig:concentration_waves}. 

To assess SRO, we examine the so-called two point correlation function, 
\begin{equation}
\Psi_{i\alpha;j\alpha'} \equiv \langle \xi_{i\alpha} \xi_{j\alpha'} \rangle - \langle \xi_{i\alpha} \rangle \langle \xi_{j\alpha'} \rangle,
\end{equation}
which is non-zero except in the high-temperature limit and can be directly related to the second derivative of the system's free energy. It tells us about the dominant atom-atom correlations in the system above any compositional order-disorder transition temperature.

\subsection{Linear Response}
A full derivation and discussion of the linear response theory for multicomponent alloys we use is given in references \cite{khan_statistical_2016, woodgate_compositional_2022}. Here we provide a brief outline of the formalism. Within the coherent potential approximation (CPA)~\cite{soven_coherent-potential_1967}, we {  start with an expression for a mean field approximation to the free energy of a system with an inhomogeneous concentration distribution, $\{\Bar{c}_{i\alpha}\}$,}
\begin{align}
    \Omega^{(1)}[\{\nu_{i\alpha}\}, \{\Bar{c}_{i\alpha}\}] =& -\frac{1}{\beta} \sum_{i\alpha} \Bar{c}_{i\alpha} \ln \Bar{c}_{i\alpha} \nonumber \\ &- \sum_{i\alpha} \nu_{i\alpha} \Bar{c}_{i\alpha} + \langle \Omega_\text{el} \rangle_0 [\{\Bar{c}_{i\alpha}\}],
\end{align}
where $\Bar{c}_{i\alpha}$ represents the mean-field average concentration of species $\alpha$ on site $i$.
The first term represents the so-called `point entropy', or `entropy of mixing'\cite{chaikin_principles_1995}. {  Each $\nu_{i\alpha}$ in the second term represents the local Lagrange parameter, specifying the concentration $\Bar{c}_{i\alpha}$ on a given site.} The final term denotes the average value of the electronic and nuclear contribution to the free energy, {  formulated within Density Functional Theory,} where the average is taken with respect to the ensemble generated by the mean-field Hamiltonian {  and consistent with the inhomogeneous concentration distribution, $\{\Bar{c}_{i\alpha}\}$}. We then expand the free energy of the system around { a homogeneous reference state, i.e the disordered solid solution},{ $\{\Bar{c}_{i\alpha}= c_\alpha\}$},  writing
\begin{align}
    \Omega^{(1)}(\{\Bar{c}_{i\alpha}\}) &= \Omega^{(1)}(\{c_{\alpha}\}) + \sum_{i\alpha} \frac{\partial \Omega^{(1)}}{\partial \Bar{c}_{i\alpha}} \Big\vert_{\{c_{\alpha}\}} \Delta \Bar{c}_{i\alpha} \nonumber \\ 
    &+ \frac{1}{2} \sum_{i\alpha; j\alpha'} \frac{\partial^2 \Omega^{(1)}}{\partial \Bar{c}_{i\alpha} \partial \Bar{c}_{j\alpha'}} \Big\vert_{\{c_{\alpha}\}} \Delta \Bar{c}_{i\alpha}\Delta \Bar{c}_{j\alpha'} + \dots.
\label{eq:landau}   
\end{align} 
The symmetry of the high-temperature, homogeneous state - the solid solution - and the requirement that any imposed fluctuation conserves the overall concentrations of each chemical species, ensures that the first-order term vanishes. {  We also set derivatives involving the on-site chemical potentials to zero as their variation is not important to the underlying physics \cite{khan_statistical_2016}.} Therefore, to second-order, the change in free energy due to a fluctuation $\{ \Delta \Bar{c}_{i\alpha} \}$ is written
\begin{equation}
    \delta \Omega^{(1)} = \frac{1}{2} \sum_{i\alpha; j\alpha'} \Delta \Bar{c}_{i\alpha} [\beta^{-1} \, C_{\alpha\alpha'}^{-1} - S^{(2)}_{i\alpha, j\alpha'}] \Delta \Bar{c}_{j\alpha'},
    \label{eq:chemical_stability_real}
\end{equation}
where $C_{\alpha \alpha'}^{-1} = \frac{\delta_{\alpha \alpha'}}{c_\alpha}$ is associated with the entropic contributions. The key quantity here is the second-order concentration derivative of the average energy of the disordered alloy, $-\frac{\partial^2 \langle \Omega_\text{el} \rangle_0}{\partial \Bar{c}_{i\alpha} \partial \Bar{c}_{j\alpha'}} \equiv S^{(2)}_{i\alpha;j\alpha'}$, related directly to the two-point correlation function. It is this quantity that our linear response theory evaluates.

{  As discussed in references \citenum{khan_statistical_2016} and \citenum{woodgate_compositional_2022}, this linear response theory accounts carefully for DFT charge density perturbation effects. This is similar to the consideration given in Density Functional Perturbation Theory (DFPT)~\cite{savrasov_linear-response_1996, baroni_phonons_2001}, used to describe lattice dynamics {\it ab initio} and response functions for phonons etc. Formally the free energy $\langle \Omega_\text{el} \rangle_0 [\{\Bar{c}_{i\alpha}\}] $ for a specific inhomogeneous concentration distribution, $\{\Bar{c}_{i\alpha}\}$, is prescribed by a DFT minimization with respect to the appropriately averaged charge and magnetization densities. There  is consequently an interdependence of the changes to the atomic occupation of the lattice sites, registered by the $\{\Delta \Bar{c}_{i\alpha}\}$, and the changes to the lattice site-resolved charge and magnetization densities which leads to a set of coupled equations~\cite{khan_statistical_2016} from which the two-point correlation function is determined. In practice these equations are solved by taking advantage of the translational symmetry of the lattice-based system and a Fourier transform to produce the correlation function $S^{(2)}_{\alpha\alpha'}(\mathbf{k})$ in reciprocal wavevector space and the change in free energy is written}
\begin{align}
    \delta \Omega^{(1)} =& \frac{1}{2} \sum_{\bf k} \sum_{\alpha, \alpha'} \Delta \Bar{c}_\alpha({\bf k}) [\beta^{-1} C^{-1}_{\alpha \alpha'} -S^{(2)}_{\alpha \alpha'}({\bf k})] \Delta \Bar{c}_{\alpha'}({\bf k})\\
    =& \frac{1}{2} \sum_{\bf k} \sum_{\alpha, \alpha'} \Delta \Bar{c}_\alpha({\bf k}) [\beta^{-1}\Bar{\Psi}^{-1}_{\alpha \alpha'}(\mathbf{k})] \Delta \Bar{c}_{\alpha'}({\bf k}).
\label{eq:chemical_stability_reciprocal}
\end{align}
The matrix in square brackets we refer to as the chemical stability matrix and is related to an estimate of the SRO, $\Psi_{i\alpha;j\alpha'}$. As we consider decreasing temperature, we look for when the lowest lying eigenvalue of this matrix, for any $\mathbf{k}$-vector in the irreducible Brillouin Zone (IBZ), passes through zero. We infer an order-disorder transition at that temperature $T_\text{us}$ with mode $\mathbf{k}_\text{us}$ and chemical polarisation { $\Delta c_\alpha$} given by the associated eigenvector.

\subsection{Effective Pairwise Interactions}

Having obtained $S^{(2)}_{\alpha \alpha'}({\bf k})$, it is then possible to fit to a real-space pairwise interaction suitable for on-lattice atomistic modelling. The Hamiltonian for our system has then the conventional Bragg-Williams~\cite{bragg_effect_1934, bragg_effect_1935} form, written:
\begin{equation}
    H = \frac{1}{2}\sum_{i \alpha; j\alpha'} V_{i\alpha; j\alpha'} \xi_{i \alpha} \xi_{j \alpha'} + \sum_{i\alpha} \nu_{\alpha} \xi_{i\alpha}, 
    \label{eq:b-w}
\end{equation}
where the $\nu_\alpha$s are chemical potentials.
For such a model $V_{i\alpha; j\alpha'}$ is equivalent to  $-S^{(2)}_{i\alpha;j\alpha'}$. {  From our linear response theory~\cite{khan_statistical_2016} outlined in section~\ref{sec:theory}, the direct correlation functions} $S^{(2)}_{i\alpha;j\alpha'}$ are calculated in reciprocal space and the $V_{i\alpha; j\alpha'}$ are recovered from them by fitting to a real-space interaction.  It should be emphasised that the earlier instability analysis is only rigorous for second-order transitions and highlight the dominant atom-atom correlations, but the $V_{i\alpha; j\alpha'}$ can be used to infer transitions which are first-order.  With this mapping, we have atom-atom interchange parameters that can be used for modelling at any temperature. {  This step assumes that the $V_{i\alpha; j\alpha'}$ calculated for the disordered solid solution (high-$T$, homogeneous limit) are the same as for low-$T$ states with order developing.} The procedure by which these pairwise interactions are obtained, via analysis of the free energy cost of compositional fluctuations around the disordered phase, makes them an unbiased best choice, being unrelated to fits to energies of specific configurations.

In practice, to obtain an effective pairwise interaction in real space, we sample a number of $\mathbf{k}$ points distributed in the irreducible wedge of the first Brillouin zone, including along lines linking the special points \cite{khachaturyan_ordering_1978}. We then fit to a function of the form
\begin{equation}
    S^{(2)}_{\alpha \alpha'}(\mathbf{k}) \approx - \sum_{n=0}^N V^{(n)}_{\alpha \alpha'} \left( \sum_{\left\{ R_i \right\}_n} e^{i \mathbf{k} \cdot \mathbf{R}_i}\right),
\end{equation}
where $\left\{ R_i \right\}_n$ denotes the set of vectors pointing to all lattice sites on the $n$th neighbour shell, and $N$ denotes the maximum number of shells considered. The $V^{(n)}_{\alpha \alpha'}$ are the coefficients fitted. { It is important to ensure that enough $\mathbf{k}$ points are sampled and enough lattice shells are included to obtain a fit which is well-converged. Typically fewer than 100 $\mathbf{k}$ points in the irreducible section of the Brillouin Zone are required to fit interactions up to the first 10 coordination shells.}

{  It is appropriate at this point to mention an important difference between the direct correlation function approach discussed here, using the $S^{(2)}_{\alpha \alpha'}(\mathbf{k})$ quantities, and related methods such as the Generalized Perturbation Method (GPM)
\cite{ducastelle_generalized_1976, gonis_configurational_1987, turchi_first-principles_1988}. The key difference is that the GPM approach makes an approximation that the charge densities on sites occupied by the different atomic species remain unchanged from those for the homogeneously disordered alloy when the concentration distribution becomes inhomogeneous. It therefore does not include the full DFT perturbation effect on the electronic density (i.e. effects of charge-transfer and charge-response)  which our approach does address. The screened GPM \cite{ruban_atomic_2004} includes part of the effect via its calculation of an electrostatic contribution to the SGPM potential. These approximations nonetheless enable effective interactions to be calculated directly in real space.}

\subsection{Monte Carlo simulations}

To explore the phase behaviour of these systems with this atomistic model, we use the Metropolis Monte-Carlo algorithm with {  with only swaps permitted to conserve overall concentrations of each chemical species, known as Kawasaki dynamics}\cite{landau_guide_2014}. This method has been used with success to study the physics of alloy formation \cite{binder_monte_1981, santodonato_predictive_2018}.

We initialise the occupation numbers at random, with the only restriction being the overall number of atoms of each species to set the concentrations. A pair of sites (not necessarily nearest neighbours) are selected at random, and the change in energy $\Delta H$ from swapping the site occupancies is calculated. If the change in energy is negative the move is accepted unconditionally, while if the change is positive the swap is accepted with probability $e^{-\beta \Delta H}$. It is important to make sure that the system is properly equilibrated at a given temperature. Our implementation applies periodic boundary conditions in all three directions. To measure the configurational contribution to the specific heat capacity (SHC) of the system, we use the fluctuation-dissipation theorem\cite{allen_computer_2017}. In equilibrium, an estimation of the specific heat is given by
\begin{equation}
    C = \frac{1}{k_b T^2} \left( \langle E^2 \rangle - \langle E \rangle^2 \right),
\end{equation}
and it is this which we calculate to obtain our SHC curves.

To quantify SRO in our simulations, we generate the Warren-Cowley SRO parameters \cite{cowley_approximate_1950, cowley_short-range_1965} adapted to the multicomponent setting, 
\begin{equation}
    \alpha^{pq}_n = 1 - \frac{P^{pq}_n}{c_q},
\end{equation}
where $n$ refers to the $n$th coordination shell, $P^{pq}_n$ is the conditional probability of an atom of type $q$ neighboring an atom of type $p$ on shell $n$, and $c_q$ is the overall concentration of atom type $q$. When $\alpha^{pq}_n>0$, $p$-$q$ pairs are disfavored on shell $n$, while when $\alpha^{pq}_n<0$ they are favored. The value 0 corresponds to the ideal, random, solid solution.

\section{Results and Discussion}
\label{sec:results}

\begin{figure*}[t]
\centering
\includegraphics[width=\textwidth]{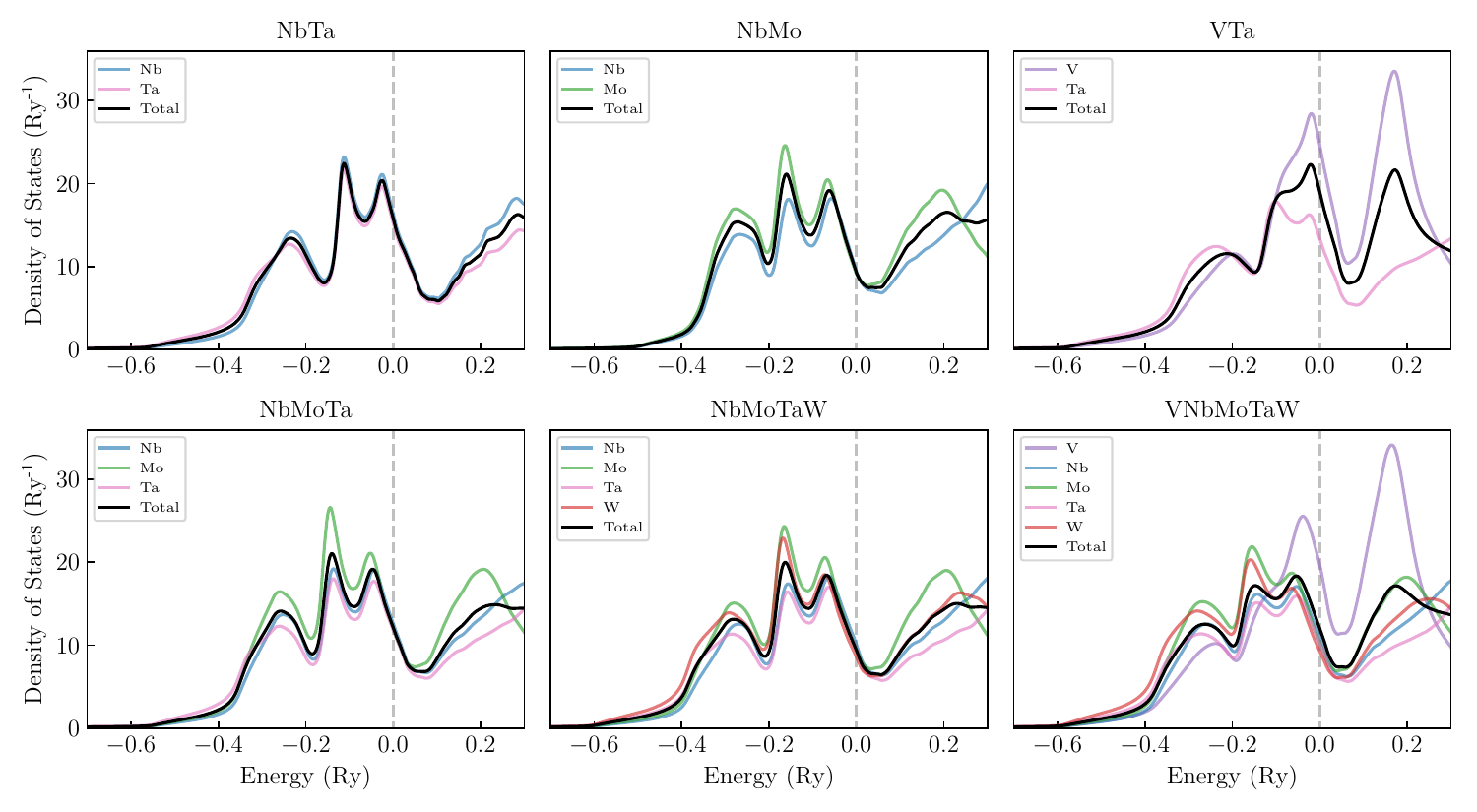}
\caption{Plots of the total and species-resolved density of states for NbTa, NbMo, VTa, NbMoTa, NbMoTaW, and VNbMoTaW around the Fermi energy. (The Fermi energy for each system is denoted by a grey, dashed, vertical line.) The total DoS curve is given by the average of the components from the separate species. NbTa is selected as an example of a $4d$/$5d$ binary where the two components are isolectronic. NbMo is selected as an example of a $4d$/$4d$ system where there is a valence difference. VTa is selected as an example of a $3d$/$5d$ system where there is an atomic size and bandwidth difference. It can be seen that, as for the binaries, in the multicomponent systems, isoelectronic species have DoS curves lying almost on top of one another, while species where there is a valence difference are separated. Being a $3d$ transition metal, the curves for V have a different profile those of the $4d$ and $5d$ elements and show the $3d$-$4/5d$ hybridized bonding states at the lower energies.}
\label{fig:dos_comparison}
\end{figure*}

\subsection{Electronic Structure Calculations}
\label{sec:electronic_structure}

\begin{table}
\centering
\begin{ruledtabular}
\begin{tabular}{lr}
Material & $a_\text{DFT}$ (\AA) \\ \hline
VTa      & 3.105 \\
NbMo     & 3.147 \\ 
NbTa     & 3.226 \\
NbMoTa   & 3.173 \\ 
NbMoTaW  & 3.151 \\ 
VNbMoTaW & 3.119 \\ 
\end{tabular}
\end{ruledtabular}
\caption{Hutsepot optimised lattice parameters for the considered alloys.}\label{table:alloy_a}%
\end{table}

To model the electronic `glue' bonding atoms together and driving SRO, we first generate the self-consistent, single-electron potentials of density functional theory (DFT) \cite{martin_electronic_2004}, which are used as the basis for performing linear response calculations. The potentials are generated in the Korringa-Kohn-Rostoker (KKR) formulation of DFT, using the coherent potential approximation (CPA) to produce an effective medium reflecting the average electronic structure of the high-temperature, high-symmetry, disordered solid solution \cite{faulkner_calculating_1980, faulkner_multiple_2018, johnson_total-energy_1990}. We use the all-electron HUTSEPOT code \cite{hoffmann_magnetic_2020} to generate these potentials although, in principle, any KKR-CPA code would also be suitable. We perform scalar-relativistic calculations within the atomic sphere approximation (ASA)\cite{stocks_complete_1978} with an angular momentum cutoff of $l_\text{max} = 3$ for basis set expansions, a $20\times20\times20$ Monkhorst-Pack grid\cite{monkhorst_special_1976} for integrals over the Brillouin zone, and a 24 point semi-circular Gauss-Legendre grid in the complex plane to integrate over valence energies. We use the local density approximation (LDA) and the exchange-correlation functional is that of Perdew-Wang\cite{perdew_accurate_1992}.

\begin{figure*}[t]
\centering
\includegraphics[width=\textwidth]{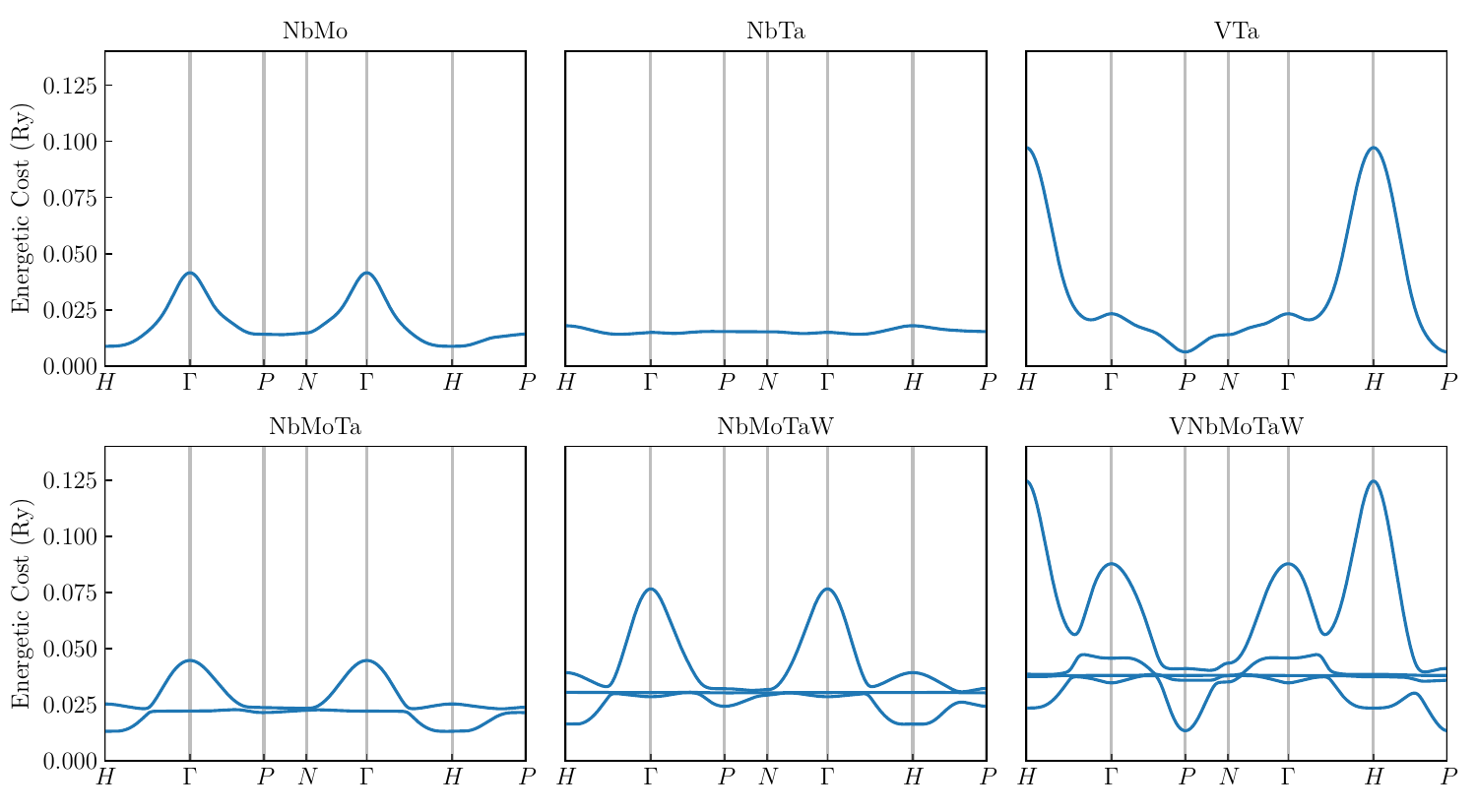}
\caption{Plots of the eigenvalues of the chemical stability matrix around the IBZ for  NbTa, NbMo, VTa, NbMoTa, NbMoTaW, and VNbMoTaW at $T=1200$K. The isoelectronic binary, NbTa, has one very flat mode, associated with weak ordering tendencies. The NbMo system, in which there is a valence difference, has a mode which dips at $H$ and rises at $\Gamma$, indicative of B2-like ordering. The VTa system is dominated by atomic size and bandwidth differences and its mode dips at $P$ and peaks at $H$, indicative of B32-like ordering. The modes present in the multicomponent systems can be interpreted as combinations of these behaviors. Flat modes are associated with correlations between isoelectronic species, modes dipping at $H$ are associated with valence differences and associated with B2-like ordering, while modes dipping at $P$ are V-dominated and associated with B32-like ordering.}
\label{fig:eigenvalue_comparison}
\end{figure*}

\begin{table*}
\begin{ruledtabular}
\centering
\begin{tabular}{l|rcrrrrr}
Material & $T_\text{us} (K) $ & $\mathbf{k}_\text{us}$ & $\Delta c_1$ & $\Delta c_2$ & $\Delta c_3$ & $\Delta c_4$ & $\Delta c_5$ \\ \hline
VTa      &    691 & $\left\{\frac{1}{2}, \frac{1}{2}, \frac{1}{2}\right\}$ &  $0.707$ & $-0.707$ &&&                      \\ 
NbMo     &    513 & $\left\{0, 0, 1\right\}$       &  $0.707$ & $-0.707$ &&&                      \\ 
NbTa     &     86 & $\left\{0, 0, 0.4\right\}$     &  $0.707$ & $-0.707$ &&&                      \\
NbMoTa   &    511 & $\left\{0, 0, 1\right\}$       &  $-0.406$ & $0.816$ & $-0.410$ &&                \\ 
NbMoTaW  &    559 & $\left\{0, 0, 1\right\}$       &  $-0.383$ & $0.594$ & $-0.595$ & $0.383$ &         \\ 
VNbMoTaW &    742 & $\left\{\frac{1}{2}, \frac{1}{2}, \frac{1}{2}\right\}$ &  $-0.824$ & $0.012$ & $0.085$ & $0.252$ & $0.500$        \\
\end{tabular}
\end{ruledtabular}
\caption{Transition temperatures, modes, and chemical polarisations for the six considered systems. {  The chemical polarisation is the eigenvector in composition space associated with the eigenvalue passing through zero. The numbering of components is indicated by the composition in the left hand column, e.g. Nb=1, Mo=2, Ta=3, W=4 in NbMoTaW, while
V=1, Nb=2, Mo=3, Ta=4, W=5 in VNbMoTaW.} As an example, in NbMoTaW, the B2 ordering {  is indicated by $\mathbf{k}_\text{us}=\{0,0,1\}$, and the chemical polarisation indicates that it} comprises one cubic sublattice rich in Nb and Ta, with the other sublattice rich in Mo and W.}\label{table:linear_response}%
\end{table*}

Along with NbMoTa, NbMoTaW, and VNbMoTaW, we study three equiatomic binary subsystems. NbTa is selected as an example of a $4d$/$5d$ binary where the two components are isoelectronic. NbMo is selected as an example of a $4d$/$4d$ system where there is a valence difference. VTa is selected as an example of a $3d$/$5d$ system where there is an atomic size and $d$-bandwidth difference. For completeness, we studied the other seven possible binary subsystems, for which the relevant results are included in the supplementary material\cite{supplementary}.

We obtain lattice parameters for all considered systems {\it ab initio}, finding the value of the lattice parameter (and therefore cell volume) for which the total DFT energy is minimised. HUTSEPOT-optimised lattice parameters for the six considered systems are tabulated in \ref{table:alloy_a}. We expect there to be a small, but systematic underestimation of lattice parameter when compared to experimental values, a well-known feature of calculations using the LDA.

For the optimised lattice parameters, we show the electronic densities of states (DoS) for the NbMoTa, NbMoTaW, and VNbMoTaW alloys, along with the three selected binaries in Fig.\ref{fig:dos_comparison}.
The distinguishing features in the DoS in these systems arise from the partially filled $d$-electron states.  V, being a $3d$ transition metal, has by far the narrowest $d$-band. Nb and Mo as $4d$ elements have narrower $d$-bands than Ta and W, but the width discrepancy is far smaller. We expect the $3d$-$4d/5d$ bandwidth difference to impact ordering in a similar manner to the Ni-Pt system, where ordering is dominated by the difference in atomic size between Ni and Pt\cite{pinski_origins_1991}. This is manifested by  $3d$-$4d/5d$-hybridized bonding states forming at lower energies and effective charge transfer to the smaller $3d$ atoms from the larger $4d/5d$ atoms. A comment should also be made about the valence of these systems. $4d/5d$ pairs such as Nb/Ta and Mo/W are isoelectronic (same valence) so their $d$-bands will naturally lie close to one another, which we expect to lead to these elements interacting weakly. Where there is a valence difference interactions are likely to be stronger.

\begin{table*}
\scriptsize
\centering
\begin{ruledtabular}
\begin{tabular}{lrrrrr|lrrrrr}
$V^{(1)}_{\alpha \beta}$ & Nb       & Mo       & Ta       &    &   & $V^{(2)}_{\alpha \beta}$ & Nb & Mo & Ta &    &   \\ \hline
Nb                       &  0.173   & $-$0.594 &  0.423   &    &   & Nb                       & $ 0.139$ & $-0.173$ & $ 0.034$ &    &   \\
Mo                       & $-$0.594 &  1.354   & $-$0.765 &    &   & Mo                       & $-0.173$ & $ 0.279$ & $-0.106$ &    &   \\
Ta                       &  0.423   & $-$0.765 &  0.343   &    &   & Ta                       & $ 0.034$ & $-0.106$ & $ 0.072$ &    &   \\
$V^{(3)}_{\alpha \beta}$ & Nb & Mo & Ta &    &   & $V^{(4)}_{\alpha \beta}$ & Nb & Mo & Ta &    &   \\ \hline
Nb                       & $-0.010$ & $-0.004$ & $ 0.014$ &    &   & Nb                       & $-0.008$ & $ 0.008$ & $-0.001$ &    &   \\
Mo                       & $-0.004$ & $ 0.117$ & $-0.114$ &    &   & Mo                       & $ 0.008$ & $-0.012$ & $ 0.004$ &    &   \\
Ta                       & $ 0.014$ & $-0.114$ & $ 0.101$ &    &   & Ta                       & $-0.001$ & $ 0.004$ & $-0.003$ &    &   \\
                         &    &    &    &    &   &                          &    &    &    &    &   \\
$V^{(1)}_{\alpha \beta}$ & Nb & Mo & Ta & W  &   & $V^{(2)}_{\alpha \beta}$ & Nb & Mo & Ta & W  &   \\ \hline
Nb                       & $ 0.509$ & $-0.938$ & $ 0.931$ & $-0.501$ &   & Nb                       & $ 0.539$ & $-0.208$ & $ 0.278$ & $-0.608$ &   \\
Mo                       & $-0.938$ & $ 0.926$ & $-0.998$ & $ 1.010$ &   & Mo                       & $-0.208$ & $ 0.274$ & $-0.225$ & $ 0.159$ &   \\
Ta                       & $ 0.931$ & $-0.998$ & $ 1.050$ & $-0.982$ &   & Ta                       & $ 0.278$ & $-0.225$ & $ 0.205$ & $-0.258$ &   \\
W                        & $-0.501$ & $ 1.010$ & $-0.982$ & $ 0.473$ &   & W                        & $-0.608$ & $ 0.159$ & $-0.258$ & $ 0.708$ &   \\
$V^{(3)}_{\alpha \beta}$ & Nb & Mo & Ta & W  &   & $V^{(4)}_{\alpha \beta}$ & Nb & Mo & Ta & W  &   \\ \hline
Nb                       & $ 0.008$ & $-0.085$ & $ 0.078$ & $-0.000$ &   & Nb                       & $-0.002$ & $-0.036$ & $ 0.025$ & $ 0.013$ &   \\
Mo                       & $-0.085$ & $ 0.176$ & $-0.173$ & $ 0.083$ &   & Mo                       & $-0.036$ & $ 0.034$ & $-0.033$ & $ 0.035$ &   \\
Ta                       & $ 0.078$ & $-0.173$ & $ 0.172$ & $-0.076$ &   & Ta                       &  $ 0.025$ & $-0.033$ & $ 0.030$ & $-0.021$ &   \\
W                        & $-0.000$ & $ 0.083$ & $-0.076$ & $-0.007$ &   & W                        &  $ 0.013$ & $ 0.035$ & $-0.021$ & $-0.026$ &   \\
                         &    &    &    &    &   &                          &    &    &    &    &   \\
$V^{(1)}_{\alpha \beta}$ & V  & Nb & Mo & Ta & W & $V^{(2)}_{\alpha \beta}$ & V  & Nb & Mo & Ta & W \\ \hline
V                        & $-1.330$ & $ 0.319$ & $-1.253$ & $ 1.681$ & $ 0.583$ & V                        & $ 4.105$ & $-0.201$ & $-0.441$ & $-1.225$ & $-2.237$   \\
Nb                       & $ 0.319$ & $ 0.350$ & $-0.633$ & $ 0.536$ & $-0.572$ & Nb                       & $-0.201$ & $ 0.302$ & $-0.296$ & $ 0.381$ & $-0.186$   \\
Mo                       & $-1.253$ & $-0.633$ & $ 1.558$ & $-0.978$ & $ 1.305$ & Mo                       & $-0.441$ & $-0.296$ & $ 0.300$ & $-0.127$ & $ 0.565$   \\
Ta                       & $ 1.681$ & $ 0.536$ & $-0.978$ & $ 0.279$ & $-1.517$ & Ta                       & $-1.225$ & $ 0.381$ & $-0.127$ & $ 0.627$ & $ 0.344$   \\
W                        & $ 0.583$ & $-0.572$ & $ 1.305$ & $-1.517$ & $ 0.202$ & W                        & $-2.237$ & $-0.186$ & $ 0.565$ & $ 0.344$ & $ 1.514$   \\
$V^{(3)}_{\alpha \beta}$ & V  & Nb & Mo & Ta & W & $V^{(4)}_{\alpha \beta}$ & V  & Nb & Mo & Ta & W \\ \hline
V                        & $ 0.099$ & $-0.082$ & $-0.006$ & $-0.026$ & $ 0.015$ & V                        & $-0.615$ & $ 0.023$ & $ 0.161$ & $ 0.108$ & $ 0.323$   \\
Nb                       & $-0.082$ & $ 0.033$ & $-0.057$ & $ 0.087$ & $ 0.020$ & Nb                       & $ 0.023$ & $ 0.011$ & $-0.036$ & $ 0.023$ & $-0.021$   \\
Mo                       & $-0.006$ & $-0.057$ & $ 0.192$ & $-0.192$ & $ 0.063$ & Mo                       & $ 0.161$ & $-0.036$ & $ 0.010$ & $-0.081$ & $-0.054$   \\
Ta                       & $-0.026$ & $ 0.087$ & $-0.192$ & $ 0.201$ & $-0.069$ & Ta                       & $ 0.108$ & $ 0.023$ & $-0.081$ & $ 0.035$ & $-0.084$   \\
W                        & $ 0.015$ & $ 0.020$ & $ 0.063$ & $-0.069$ & $-0.028$ & W                        & $ 0.323$ & $-0.021$ & $-0.054$ & $-0.084$ & $-0.163$  
\end{tabular}
\end{ruledtabular}
\caption{Interchange parameters for the three multicomponent systems considered, {  fitted from $S^{(2)}$s evaluated at $T=1200$K for the specified equiatomic composition}. All values in mRy. Interactions are dominated by the first two shells.}\label{table:interchange}%
\end{table*}

\subsection{Linear Response Analysis}
\label{sec:linear_response}

Starting from the self-consistent potentials and electron densities of the ideal solid solution, we use our theory to construct the chemical stability matrix in reciprocal space. In Figure \ref{fig:eigenvalue_comparison}, we plot the eigenvalues of this matrix along various symmetry lines of the irreducible Brillouin Zone (IBZ) for the six considered systems evaluated at 1200K. As for the DoS, plots for all possible equiatomic binary systems are included in the supplementary material\cite{supplementary}. Then, in Table \ref{table:linear_response}, we give our predicted ordering temperatures, associated modes, and chemical polarisations. We emphasise that these ordering temperatures are computed within a mean-field theory and are therefore expected to be overestimates of {  exact} ordering temperatures in these systems.

When the eigenvalue plots of Fig. \ref{fig:eigenvalue_comparison} are considered, we first look at the three binaries to understand the multicomponent systems. The mode present in NbMo, dipping at $H$ and peaking at $\Gamma$ is associated with a difference in valence between the two species and is indicative of B2-like ordering tendencies. The exceptionally flat mode present in NbTa is associated with very weak interactions and has its origins in the fact that Nb and Ta are isoelectronic.  Finally, for VTa, the mode which has a strong peak at $H$ and dips at $P$ can be associated with the effects of differing atomic size and bandwidth between $3d$ and $4d$/$5d$ elements and is indicative of B32-like order. 

In our calculations this appears as an effective charge transfer to the small V atoms from the bigger Ta atoms screened by the valence electrons; in general, for these multicomponent systems, our calculations find electron density to be transferred from the $3d$ transition metal atoms to the $4d$/$5d$ ones. For an equiatomic binary system on the BCC lattice, when considering first and second nearest neighbour shells, the B2 structure has 8 unlike and 6 like neighours, while the B32 structure has 10 unlike and 4 like neighbours. A screened Coulombic interaction will therefore favor the B32 structure over B2. We find this happens strongly for VW, a predicted B32-former\cite{muzyk_phase_2011}, where the size effects dominate the valence difference. We find the ordering temperature for VW to be 905K. The associated data can be found in the supplementary material\cite{supplementary}.

Modes associated with the above mechanisms can be observed in the multicomponent systems, albeit with mixing between the modes. Sample eigenvalues and associated eigenvectors of the chemical stability matrix at special points of the IBZ are tabulated in the supplementary material.

Looking at the tabulated transition temperatures and predicted ordering, we predict B2-like ordering in the ternary NbMoTa at a temperature of 511K. The polarisation in concentration space indicates one sublattice rich in Nb and Ta, with the other rich in Mo. This indicates that it is energetically favourable for nearest neighbors in this system to have a valence difference. 

Our results for NbMoTa are also consistent with the quarternary NbMoTaW system, in which we again find a B2-like ordering at 559K, with Nb and Ta segregating onto one sublattice, Mo and W on the other. This type of ordering is consistent with previous theoretical works on this system \cite{kormann_long-ranged_2017, singh_ta-nb-mo-w_2018, kormann_interplay_2016, widom_hybrid_2014, huhn_prediction_2013, singh_atomic_2015, kostiuchenko_impact_2019}, demonstrating the robustness of our approach. Our ordering temperature for this system is also consistent with some earlier works. Ref.~\citenum{kormann_interplay_2016} used the projector-augmented wave (PAW) method with chemical disorder simulated using special quasi-random structures (SQS) and included effects of vibrations, electronic excitations, and configurational entropy. A transition temperature of 717K was found without lattice relaxations, while with lattice lattice relaxations it was reduced to 508K. More recently, Ref.~\citenum{kostiuchenko_impact_2019} used interatomic potentials generated via machine learning to predict an ordering temperature to a B2 phase of approximately 600K without atomic relaxations, and approximately 300K when their effects are included.

For the quinary VNbMoTaW system, we find ordering to be dominated by V and is B32-like, with a predicted ordering temperature of 742K. This is consistent with Ref.~\citenum{fernandez-caballero_short-range_2017}, which used a cluster expansion Hamiltonian to predict B32-like order emerging at around 750K, which is also consistent with the analysis of the V-W binary system\cite{muzyk_phase_2011, supplementary}.

A comment should be made about the importance of including effects of charge-transfer and charge-response in these calculations, {  which is the principal difference of our approach from other CPA-based approaches such as the GPM as discussed in section~\ref{sec:theory}}.  If we use  the so-called ``band-only'' approximation in our linear response calculation and assume the one-electron potentials to be insensitive to the configurational environment, we obtain markedly different results for these materials. Indeed, we find that, for the five-component VNbMoTaW system, V is predicted erroneously to phase-segregate out of the solid solution at a very high temperature. For the four component NbMoTaW system, within the band-only approximation, we find that the minimum eigenvalue does not pass through zero at the $H$-point exactly, which would indicate long-ranged interactions and anomalous order as suggested in some earlier work on this system\cite{kormann_long-ranged_2017, singh_ta-nb-mo-w_2018}.  Including fully the effects of charge-transfer and charge-response remedies both of these issues; V is found to integrate well in the five-component system and simple, B2-like ordering is predicted in the four-component, with no significant long-ranged interactions. We noted that these effects were also significant in our study of compositional order in the Cantor-Wu systems \cite{woodgate_compositional_2022}, and suggest that they need to be accounted for carefully in any theory of phase stability in multicomponent alloys.

\subsection{Pairwise Interactions}

Table~\ref{table:interchange} shows $V^{(n)}_{\alpha \alpha'}$ for $n=1, 2, 3, 4$ for the three multicomponent systems. It can be seen that the interactions are dominated by first- and second-nearest neighbours, and we therefore conclude that any model limited to nearest-neighbour distance only will fail to capture the relevant physics. Interactions being strongest on the first two neighbour shells is also consistent with recent results obtained using effective pair interaction generated using machine learning on a DFT data-set\cite{liu_monte_2021}.

By far the strongest interacting element is V, consistent with both the pointers from the DoS plots and the results of the linear response analysis. Significantly, a number of interactions involving V are larger on the second shell than on the first, consistent with B32-like order.

\subsection{Atomistic Modelling}

\begin{figure}
\centering
\begin{subfigure}{\linewidth}
\begin{subfigure}{0.3\textwidth}
\centering
\includegraphics[height=\textwidth]{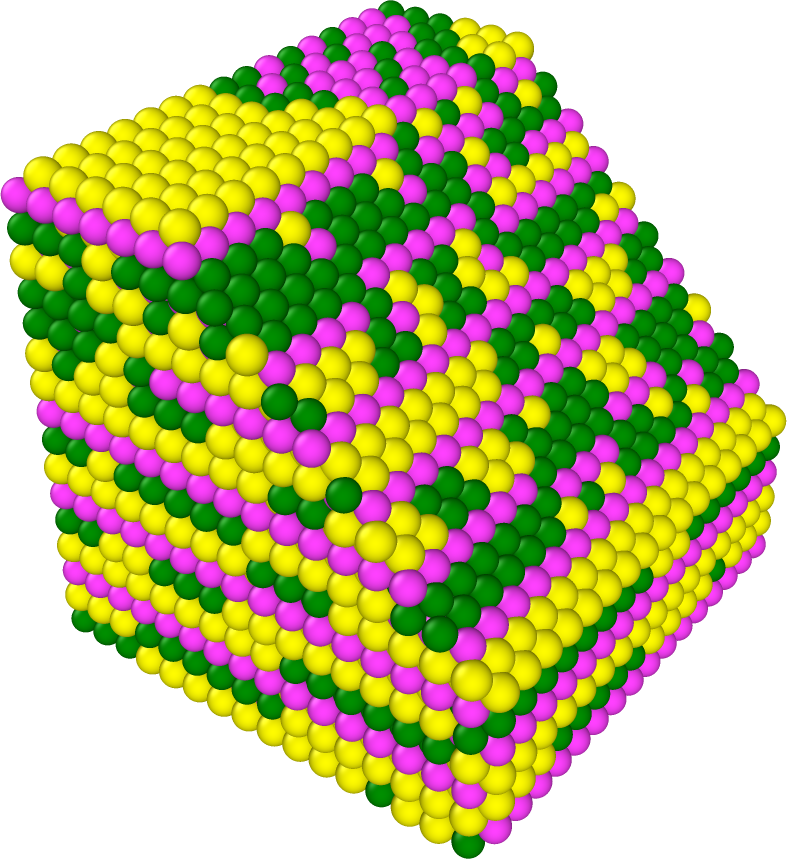}
\caption*{10K}
\end{subfigure}
\begin{subfigure}{0.3\textwidth}
\centering
\includegraphics[height=\textwidth]{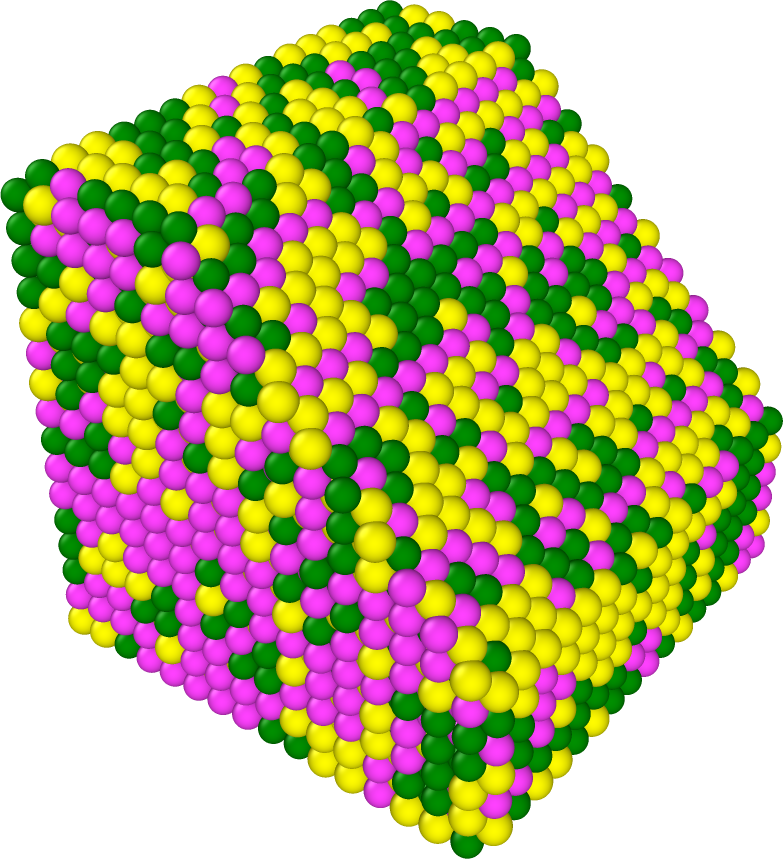}
\caption*{300K}
\end{subfigure}
\begin{subfigure}{0.3\textwidth}
\centering
\includegraphics[height=\textwidth]{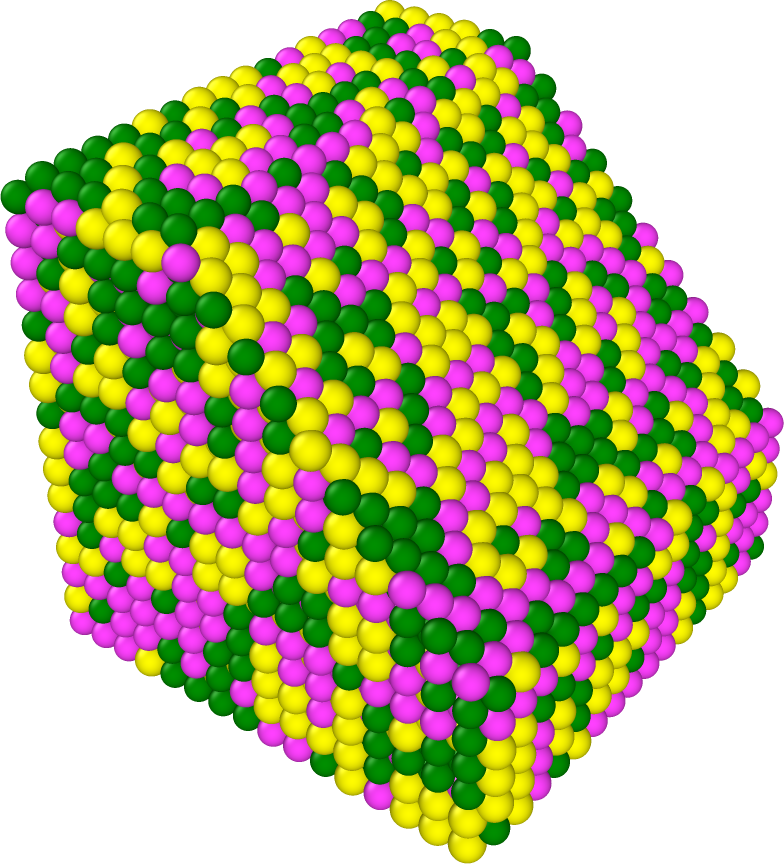}
\caption*{1200K}
\end{subfigure}
\caption{NbMoTa}
\end{subfigure} \\
\centering
\begin{subfigure}{\linewidth}
\begin{subfigure}{0.3\textwidth}
\centering
\includegraphics[height=\textwidth]{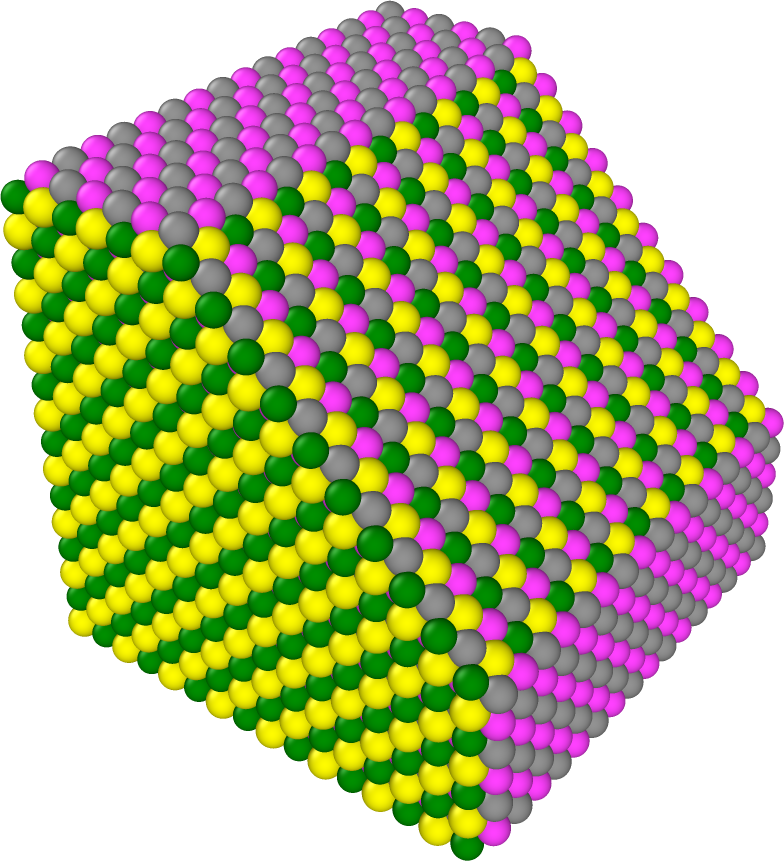}
\caption*{10K}
\end{subfigure}
\begin{subfigure}{0.3\textwidth}
\centering
\includegraphics[height=\textwidth]{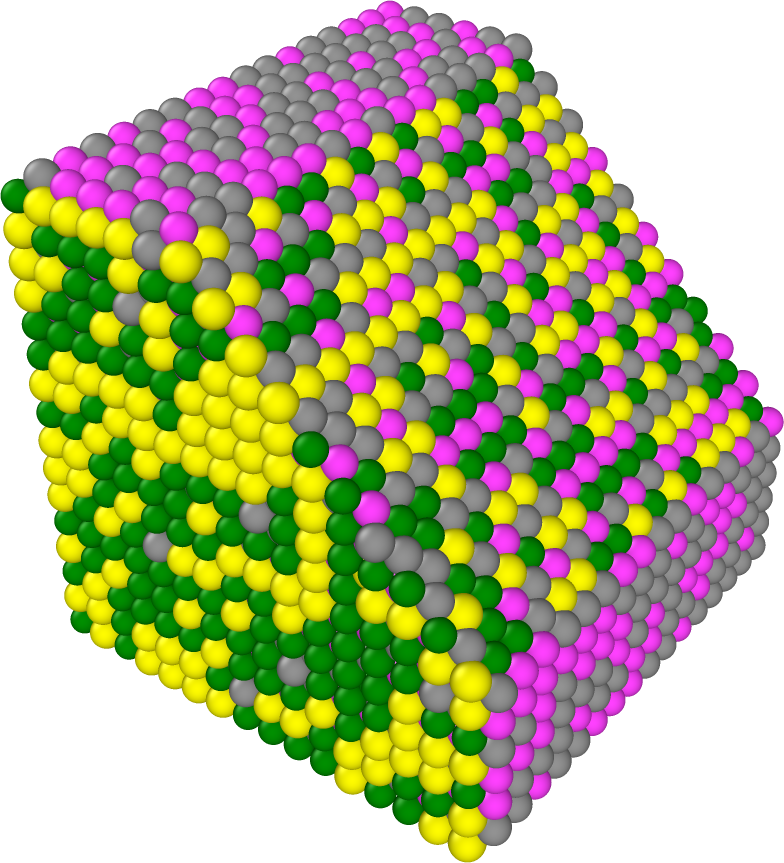}
\caption*{300K}
\end{subfigure}
\begin{subfigure}{0.3\textwidth}
\centering
\includegraphics[height=\textwidth]{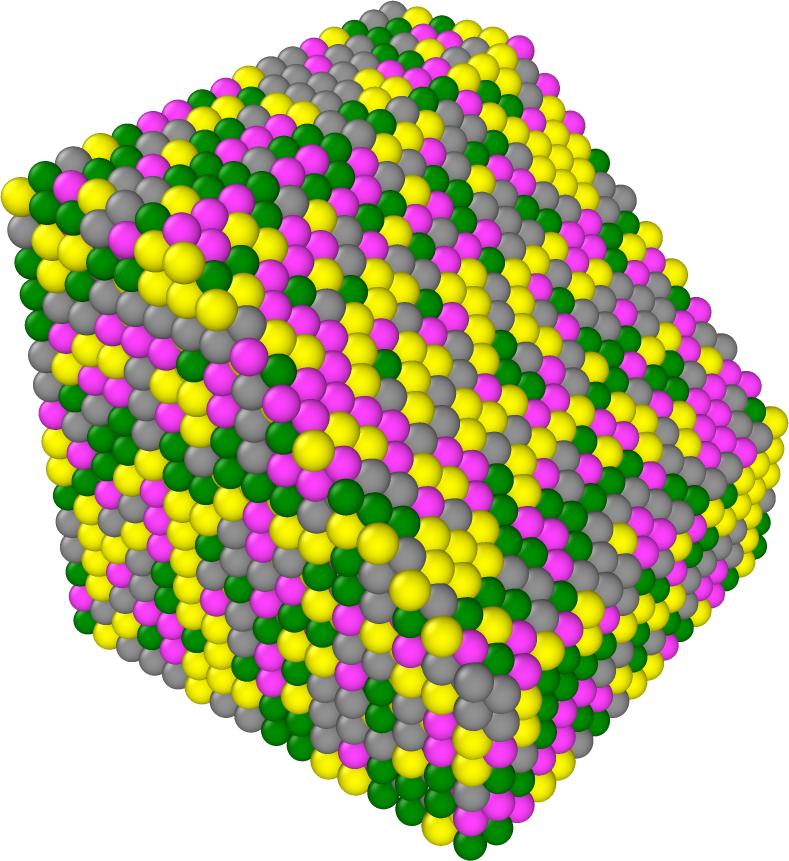}
\caption*{1200K}
\end{subfigure}
\caption{NbMoTaW}
\end{subfigure} \\
\centering
\begin{subfigure}{\linewidth}
\begin{subfigure}{0.3\textwidth}
\centering
\includegraphics[height=\textwidth]{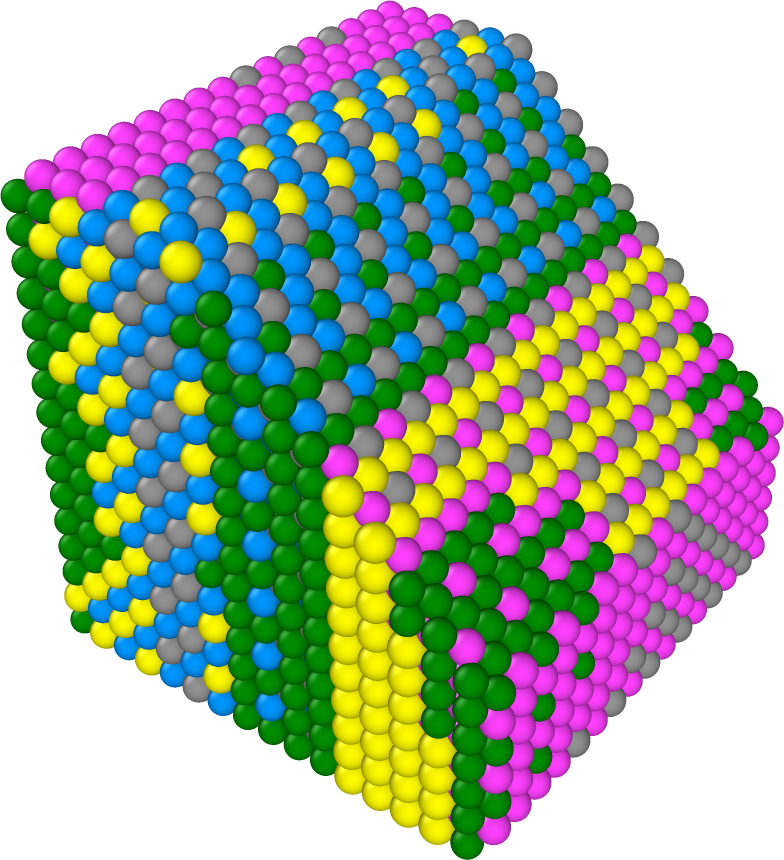}
\caption*{10K}
\end{subfigure}
\begin{subfigure}{0.3\textwidth}
\centering
\includegraphics[height=\textwidth]{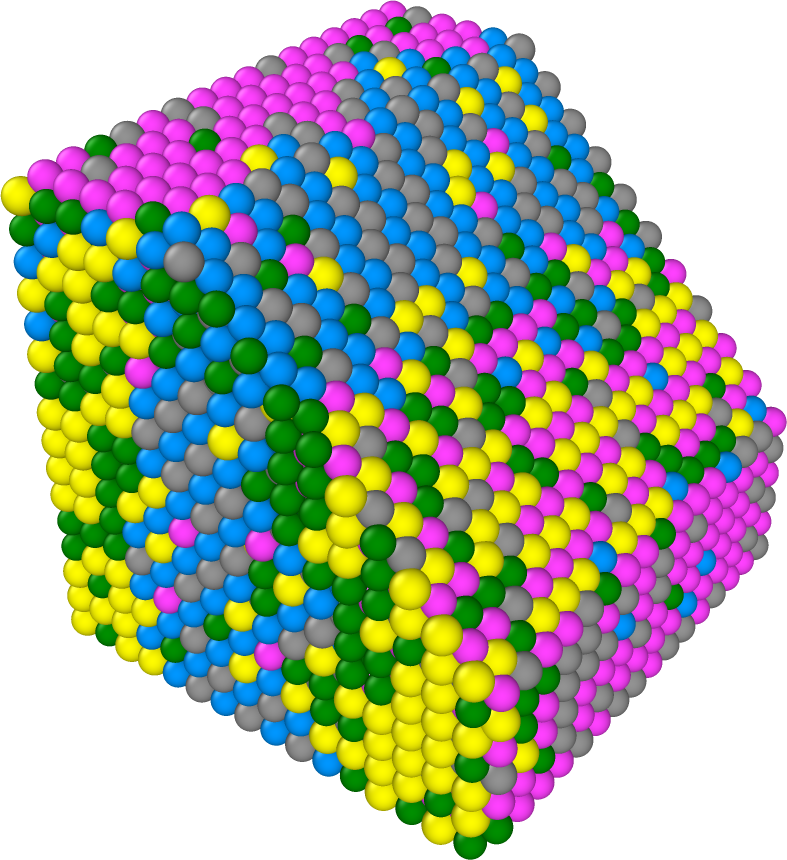}
\caption*{300K}
\end{subfigure}
\begin{subfigure}{0.3\textwidth}
\centering
\includegraphics[height=\textwidth]{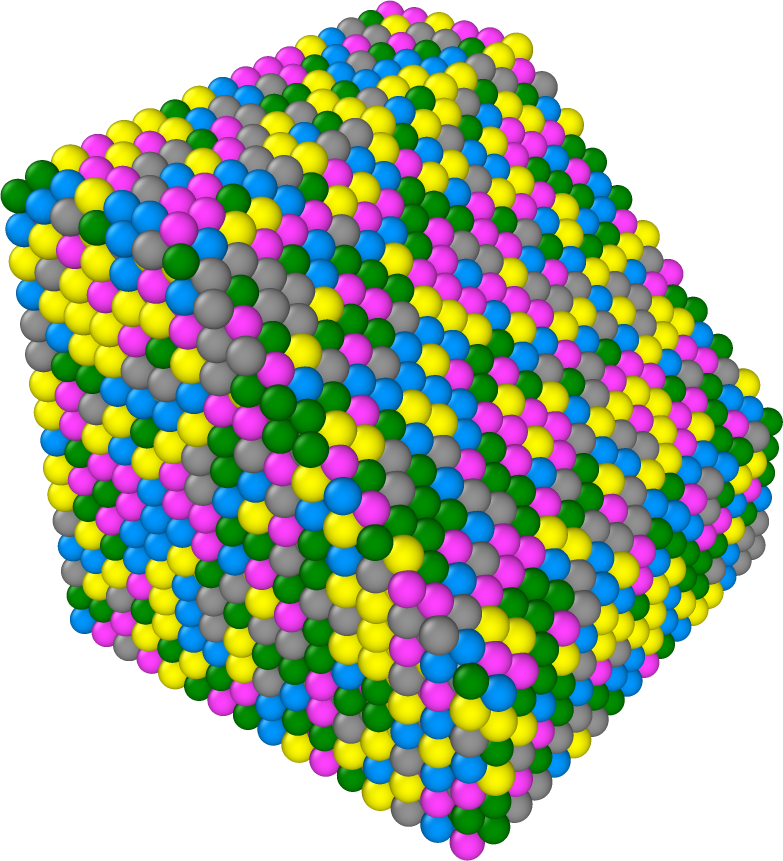}
\caption*{1200K}
\end{subfigure}
\caption{VNbMoTaW}
\end{subfigure} \\
\caption{Visualised configurations from Monte Carlo simulations for the three considered multicomponent systems at temperatures of 1200K, 300K and 10K. V, Nb, Mo, Ta, and W are coloured blue, green, pink, yellow, and grey respectively. A cut has been made through the simulation cell to make ordered structures more clearly visible. {  In the case of NbMoTaW, the emergence of a layered, B2-like structure can be seen in the $T=300$K configuration. For the ternary NbMoTa, some layering can be seen relative to the 1200K configuration, but it is less clear than for the quarternary.} For the five component system, the system demonstrates multiphase behaviour, with patches of B32-like order between V and W, and other patches of order akin to that observed in the ternary NbMoTa. Images generated using OVITO \cite{stukowski_visualization_2010}.}
\label{fig:visualisations}
\end{figure}

\begin{figure*}
\centering
\includegraphics[width=\textwidth]{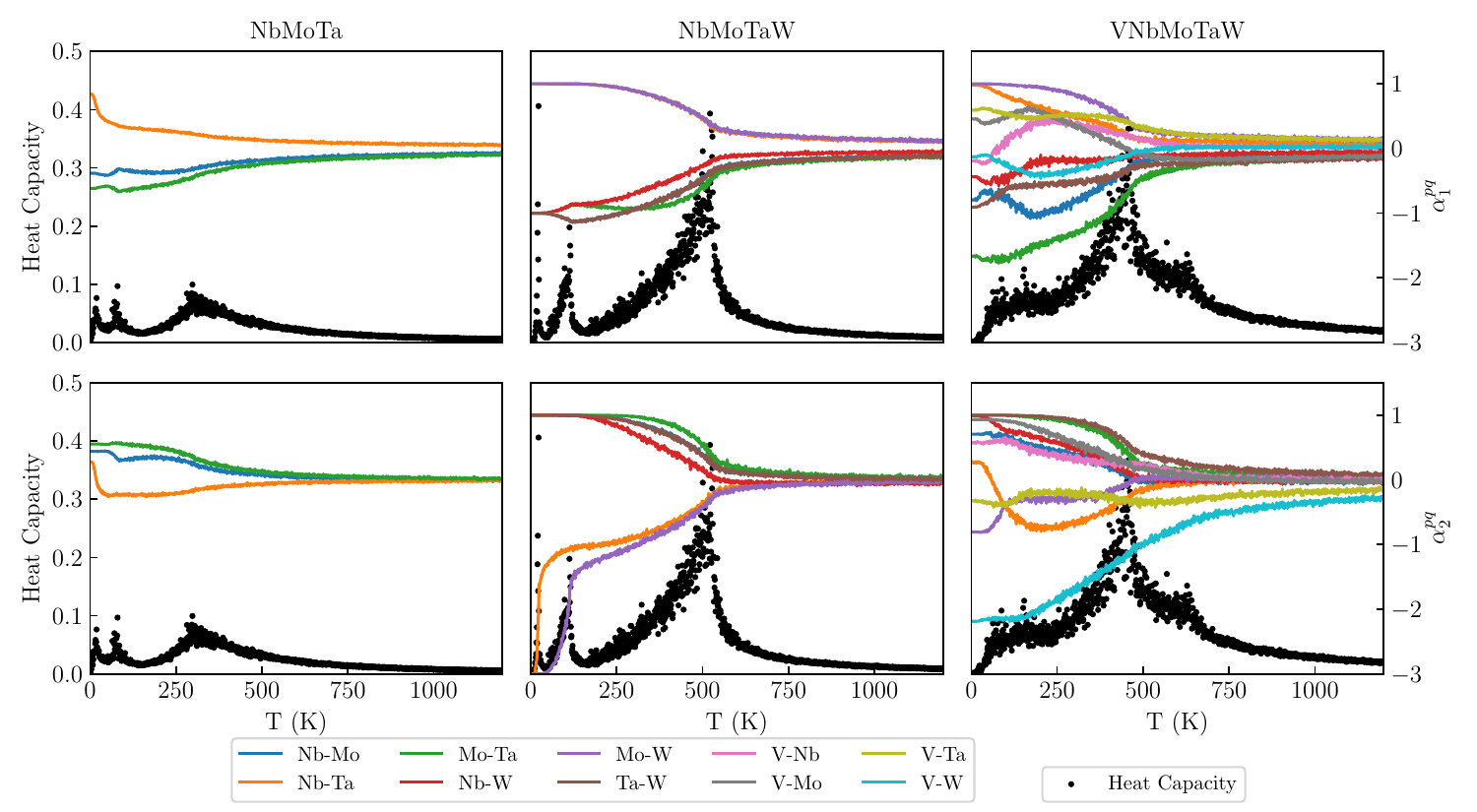}
\caption{Plots of the Warren-Cowley SRO parameters and configurational contribution to the SHC as a function of temperature for the three multicomponent systems considered from lattice-based Monte Carlo simulations using our extracted pairwise parameters. We show $\alpha^{pq}_n$ for $n=1,2$. NbMoTa shows little SRO, with the only notable feature being towards pairs with a valence difference on the first shell, and away from those pairs on the second shell, indicative of B2-like ordering. The same trend can be seen more strongly in NbMoTaW. The most notable trend on the five component plots is that V-W is largely indifferent on the first shell, but highly favoured on the second, {  a precursor to a B32-like structure.}}
\label{fig:monte_carlo}
\end{figure*}

Using the obtained pairwise interactions, we performed lattice-based Monte Carlo (MC) simulations for NbMoTa, NbMoTaW, and VNbMoTaW to better understand the nature of SRO in these systems and probe ordering below the initial order-disorder transition temperature. A lattice-based model is suitable for these systems because the BCC structure is capable of accommodating large atomic size discrepancies\cite{kube_phase_2019}.  All calculations used a $16\times16\times16$ cubic unit cells, each with 2 lattice sites per unit cell, for a total of 8192 atoms. Periodic boundary conditions were applied. The systems were prepared in an initially random configuration, then annealed from 1200K to 10K in steps of 1K, with $10^3$ MC steps per atom at each temperature.

Figure~\ref{fig:monte_carlo} shows plots of the Warren-Cowley SRO parameters and SHC curves for NbMoTa, NbMoTaW, and VNbMoTaW, while Figure~\ref{fig:visualisations} shows sample visualised configurations from our simulations.

In NbMoTa, little SRO is observed, although it is consistently seen that Nb and Ta favour Mo as a nearest neighbour, indicative of a B2-like ordering and consistent with our earlier linear response analysis. The visualised low-temperature configuration shows no clear single-phase ground state, although this is to be expected at this stoichiometry.

For the well-studied four component NbMoTaW, it can be clearly seen that Mo-W and Nb-Ta pairs (isoelectronic) are disfavored as nearest neighbours, while pairs Nb-W, Mo-Ta, Nb-Mo, and Ta-W (pairs with a valence difference) are favored, indicative of B2-like order, consistent again with our linear response analysis and also with earlier literature. This B2-like ordering is followed at lower temperatures by ordering on each of the two sublattices, resulting in a Heusler-type ground state, as can be seen in the visualised configurations. The ordering between Mo and W on one sublattice emerges earlier than between Nb and Ta in our simulations. The ground state obtained in our simulations is not the same as in some other works, notably Ref. \citenum{kostiuchenko_impact_2019}, which predicted a layered arrangement. However, our approach is most valid at high temperatures, where the pairwise parameters for atomistic modelling are extracted, so we do not view this low-temperature disagreement as an issue.

Finally, for the five component VNbMoTaW, the picture is less clear-cut. The strongest trends are towards Mo-Ta and Nb-Mo pairs at nearest neighbour distance, which could be interpreted as B2-like ordering, as for the four component. However, at second-nearest neighbour distance, V-W pairs are highly favored, suggesting instead B32-like order. When we visualise our configurations, the picture becomes clear, however, because there is clear multiphase behaviour emerging, with seperate regions of B32-like V-W and patches of B2-like order involving the other elements present, although this emerges at low temperatures and is unlikely to be experimentally observable.

\section{Conclusions}
\label{sec:conclusions}

Our results suggest that there is a set of simple underlying mechanisms driving atomic SRO in refractory HEAs. Pairs of $4d$/$5d$ elements which are isoelectronic interact weakly, mix well and make little or no contribution to SRO in a material. Pairs of $4d/5d$ elements with a valence difference are favored as nearest neighbours and drive B2-like ordering. For NbMoTa we find a B2-like (Nb, Ta; Mo) ordering, and for NbMoTaW we find a similar B2-like (Nb,Ta;Mo,W) ordering.

The addition of the $3d$ element V, with its smaller atomic size and narrower $d$-band drives a different, B32-like order, competing with and eventually dominating the B2-like state. We suggest that this small atom-big atom effect is important in understanding phase stability in many $3d$-$4d$/$5d$ multicomponent alloys. Our calculations on VNbMoTaW also give order more consistent with a B32-like structure emerging at a sufficiently high temperature that SRO in this system may be experimentally obersvable given suitable heat treatment. The results are further validated by our analysis of all possible equiatomic binary systems, given in the Supplementary Material.

Our results on SRO are consistent with earlier works on these systems, and we provide insight into the underlying physics driving ordering by studying the electronic structure of the disordered solid solution. Moreover, by using an effective medium theory, the CPA, and a simple pairwise model, we are able to obtain our results using a fraction of the computing resources taken by studies which require large numbers of DFT calculations on supercells to train potentials. All figures in this work were produced using less than 500 CPU hours on the {\it Orac} cluster at the University of Warwick, which uses Intel E5-2680 v4 (Broadwell) processors.

In conjunction with our earlier study on the Ni-based Cantor-Wu alloys, we take our results as evidence that our approach provides accurate results for very little computational cost, and is therefore an ideal candidate for searching for new HEA compositions and novel intermetallic compounds for a variety of applications. We are in the process of adapting our codes for high-throughput calculations to rapidly explore this vast space of candidate materials.

\begin{acknowledgments}
The present work formed part of the PRETAMAG project,
funded by the UK Engineering and Physical Sciences Research
Council, Grants No. EP/M028941/1, EP/W021331/1. C.D.W. is supported by a studentship within the UK Engineering and Physical Sciences Research
Council-supported Centre for Doctoral Training in Modelling of Heterogeneous Systems, Grant No. EP/S022848/1.
\end{acknowledgments}

\end{document}